\RequirePackage{fix-cm}
\documentclass[smallextended]{svjour3}       
\smartqed  
\usepackage{graphicx}
\usepackage{amssymb}
\usepackage{amsmath}
\usepackage{graphicx}
\usepackage{dcolumn}
\usepackage{bm}

\begin{document}

\title{Quantum correlation via quantum coherence\thanks{This work was supported by the National Natural
Science Foundation of China, under Grants No.11375036 and No. 11175033, and
by the Fundamental Research Funds of the Central Universities, under Grant
No. DUT12LK42.
}
}


\author{Chang-shui Yu         \and
        Yang Zhang \and Haiqing Zhao
}


\institute{Chang-shui Yu\at
              School of Physics and Optoelectronic Technology, Dalian University of
Technology, Dalian 116024, P. R. China \\
                         \email{quaninformation@sina.com; or ycs@dlut.edu.cn}             \\
\and Yang Zhang\at School of Physics and Optoelectronic Technology, Dalian University of
Technology, Dalian 116024, P. R. China \\
           \and
           Haiqing Zhao \at
              School of Science, Dalian Jiaotong University , Dalian 116028, P. R.
China
}

\date{Received: date / Accepted: date}

\maketitle

\begin{abstract}
Quantum correlation includes quantum entanglement and quantum discord. Both entanglement and discord
have a common necessary condition--------quantum coherence or quantum superposition.
In this paper, we attempt to give an alternative understanding of how quantum correlation is related to quantum coherence.
We divide the coherence of a quantum state into several classes and find  the complete coincidence
between geometric (symmetric and asymmetric) quantum discords and some particular classes of quantum coherence. We propose
a revised measure for total coherence and find that this measure can lead to a symmetric version of
geometric quantum correlation which is analytic for two qubits. In particular, this measure can also arrive at a monogamy equality on the distribution of quantum coherence. Finally, we also quantify a remaining type of
quantum coherence and find that  for two qubits it is directly connected with quantum
nonlocality.
\end{abstract}



\section{Introduction}

Quantum coherence, or quantum superposition, is one of the most fundamental features of quantum world
that are distinguished from the classical one. Associated with the tensor structure of the composite system, it
directly leads to the formation of quantum entanglement which is usually treated as another key
 quantum mechanical feature and is so important that it has been recognized to be
an important physical resource in quantum information processing (QIP). However, quantum coherence is only a necessary 
condition for the entanglement.  In order to get a good understand of quantum entanglement,
most  including ourselves have been making efforts [1-13] to find out
how to tell whether a state is entangled or not, or to what degree a given state is entangled, and to reveal the properties of entanglement measure
of different types.  Here we would like to ask the first question:  how entanglement is related to quantum coherence?

 Quantum entanglement is so familiar  to us that it could be the first candidate when quantum correlation is mentioned.
  However, quantum
entanglement does not cover all the quantumness of
correlations. It has been shown that quantum discord can effectively quantify the quantum correlation  [14-17]. It has
been shown to have some similar properties to quantum entanglement, such as
demonstrating the quantum advantage in some QIP, but it is beyond quantum
entanglement because it can be present even in separable mixed states [18].
In the past few years, quantum discord has attracted many interests in the various fields [19-30], 
but the understanding of quantum discord remains limited. For example,
the original quantum discord is an information-theoretic one that is only
analytically calculated for some special states [26,27], even though the
geometric quantum discord has the analytic expression for all two-qubit
states [31]; Quantum discord is not symmetric if we exchange the two
subsystems of the measured quantum state, which even shows completely
opposite behavior; In particular, quantum discords in terms of different
definitions are not consistent with each other for the ordering of some
quantum states [20]. Recently, several
attempts on the operational interpretation of quantum discord have been
given to the information theoretic quantum discord which is related to the
quantum state merging [32,33] and the relative-entropy-based quantum discord
which is connected with distillable entanglement [34]. Thus we come up with the second natural question: 
 How can we understand the geometric quantum discord from the most fundamental
quantum mechanical feature---------quantum coherence?

In this paper, we will answer the above two questions by studying the relation between geometric quantum discord, quantum entanglement
and  quantum coherence. We classify the quantum coherence into three classes
based on some particular approaches, whilst some measure can be naturally
given to the quantum coherence of different classes. We find that geometric
quantum discord of each possible type (including the symmetric and asymmetric version) is just consistent with a special class of quantum coherence, which serves as the answer to our second question.
We also suggest  a new measure for the total quantum coherence. It can serve as a symmetric geometric quantum
correlation and can be analytically
calculated for the systems of two quits. In particular,
the new symmetric quantum correlation for pure states is equivalent to the squared concurrence.  Associated with the 
 coherence of the subsystems, we find  an interesting monogamy equation that shows how quantum coherence is distributed or how entanglement is generated similar to 3-tangle introduced in Ref. [35].
  In addition, we quantify the
third class of quantum coherence and find that  for the systems of two qubits, this class of quantum coherence measure is directly connected
with some quantum nonlocality, i.e., the violation of Clauser-Horne-Shimony-Holt (CHSH) inequality [36].
Thus for the system of two qubits, not only quantum entanglement and quantum discord but also quantum nonlocality in terms of CHSH inequality can be understood in the fundamental frame---------quantum coherence.
Our main results are listed in detail in Sec. II which is organized as follows. We first divide quantum coherence into three class, then we prove each kind of geometric quantum 
discord is consistent with a kind of quantum coherence, and then we show that our suggesting measure for the total coherence  can serve as a symmetric quantum correlation measure, prove its equivalence to the
squared coherence for pure states and show how it is related to
the distribution of quantum coherence. By using this distribution property, we also use a constructive way to show the analytic quantum discord for pure states. Finally, we find out that the third class of quantum coherence can be related with the violation of CHSH inequality for bipartite states of qubits.

\section{Quantum coherence and quantum discord}

\subsection{Quantum coherence}

Quantum coherence arises from quantum superposition, which is a necessary
condition for quantum correlations. Generally speaking, a good definition of
quantum coherence does not only depend on the state of the system $\rho $,
but also depend on the alternatives under consideration which are usually
attached to different eigenvalues of an observable $A$. Since the
off-diagonal elements of $\rho $ characterize interference, they are usually
called \textit{coherences} with respect to the basis in which $\rho $ is
written [37,38]. The measurements on the observables that do not commute
with $A$ can reveal the interference. Based on different viewpoints, a lot
of quantum coherence measure can be defined with respect to the basis
[39,40]. Here we will present a new coherence measure by which we can
classify the coherence.
\begin{figure}[tbp]
\includegraphics[width=0.75\textwidth]{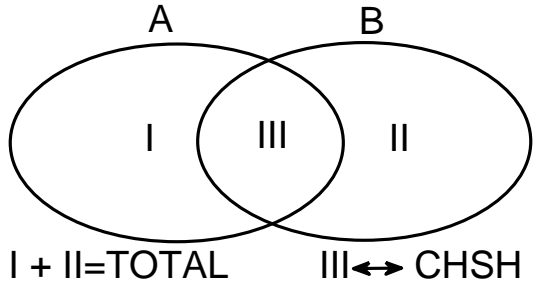}
\caption{Illustration of the classification of quantum coherence. Each
ellipse corresponds to one class of coherence. The overlap of the two
ellipses corresponds a special type of coherence. \textquotedblright
TOTAL\textquotedblright\ means the total coherence. \textquotedblright
CHSH\textquotedblright\ means that Class $III$ is connected with the
violation of CHSH inequality, which is only satisfied by two qubits.}
\label{1}
\end{figure}

In order to explicitly present our classification of quantum coherence and
the corresponding measure, we consider an $n_{1}\otimes n_{2}$ density matrix
\begin{equation}
\rho =\sum_{i,k=0}^{n_{1}-1}\sum_{j,l=0}^{n_{2}-1}\rho _{ij,kl}\left\vert
ij\right\rangle \left\langle kl\right\vert ,
\end{equation}%
with $\rho _{ij,ij}>0$, $\sum \rho _{ij,ij}=1$ and tr$\rho ^{2}\leq 1$. As
mentioned in our previous paper [41], the coherence can be measured by the
contribution of the off-diagonal entries of $\rho $. Although it provided an
explicit geometric meaning of coherence in a given basis, it is obviously
basis-vector (local basis-vector included) dependent. However, is the
characteristics of the off-diagonal entries in the coherence measure the
same? The answer is no. Now let us consider a scheme of local operations and
classical communication. Suppose Alice and Bob share a quantum state $\rho $
given in Eq. (1). If Alice performs von Neumann measurement on her qubit in
terms of the same basis of $\rho $ and then informs Bob her measurement
outcomes, Bob can obtain all the exact information about the off-diagonal
entries of $\rho $ that are related to his reduced density matrix at least in
theory (for example, based on his local quantum state tomography [42,43]).
The other off-diagonal entries of $\rho $ can not be attainable. So these
unattainable off-diagonal entries can be regarded as Class $I$. On the
contrary, if Bob performs von Neumann measurements on his qubit first, there
will also exist some entries that Alice can not attain. These Alice's
unattainable entries can be regarded as Class $II$. It is obvious that Class 
$I$ and Class $II$ have an overlap, which corresponds to the the
off-diagonal entries along the anti-diagonal line of $\rho $. The entries in
the overlap can not be attained by either Alice or Bob. We call the overlap
as Class $III$. Thus from the point of view of coherence, we have classified
the quantum coherence into three classes. Why do we classify the quantum
coherence with such a scheme? One can also understand it in a physical way.
When one subsystem of $\rho $ undergoes any a quantum channel [44], one can
find that the entries in Class $i$, $i=I,II$ have different decoherence
rates from the rest, while the decoherence of the entries in Class $III$
will happen if either subsystem undergoes a quantum channel. These can be
shown for two qubits more obviously under the phase-damping channel (pure
decoherence [45]). This classification is something like the classification
of tripartite mixed-state entanglement [46], where the different classes
have overlaps.

\subsection{Geometric quantum discord}

Now in order to measure the degree of quantum coherence, we introduce
quantum coherence measures for each class. For a given $\left( m\otimes
n\right) $-dimensional $\rho $ with different bases considered, i.e. $\rho
_{U_{A}\otimes U_{B}}=\left( U_{A}\otimes U_{B}\right) \rho \left(
U_{A}^{\dagger }\otimes U_{B}^{\dagger }\right) $. In order to collect the
contributions of Bob's unattainable off-diagonal entries (Class $I$) from $%
\rho _{U_{A}\otimes U_{B}}$, we can select the entries as 
\begin{equation}
\Delta _{kk^{\prime },ll^{\prime }}=\left\langle kk^{\prime }\right\vert
\rho _{U_{A}\otimes U_{B}}\left\vert ll^{\prime }\right\rangle ,k\neq l
\end{equation}%
where $\left\vert k\right\rangle $, $\left\vert l\right\rangle $ and $%
\left\vert k^{\prime }\right\rangle $, $\left\vert l^{\prime }\right\rangle $
are the computational basis of the subsystem A and B, respectively. So the
contribution to quantum coherence in the sense of squared $l_{2}$ norm of
matrix $[\Delta _{kk^{\prime },ll^{\prime }}]$ is given by%
\begin{equation}
C_{[A|B]}(\rho )=\sum\limits_{k^{\prime },l^{\prime },\left( k\neq l\right)
}\left\vert \Delta _{kk^{\prime },ll^{\prime }}\right\vert ^{2}.
\end{equation}%
Thus Alice and Bob can select a proper frame ($U_{A}$ and $U_{B}$) such
that $C_{[A|B]}(\rho )$ can be minimized after their operations. 
Similarly, for the  coherence in Class $II$, we have $C_{[B|A]}(\rho )$ can be given by
\begin{equation}
C_{[B|A]}(\rho )=\sum\limits_{k^{\prime }\neq l^{\prime },k,l}\left\vert
\Delta _{kk^{\prime },ll^{\prime }}\right\vert ^{2},
\end{equation}%
In this
way, we have the following definition.

\textit{Definition 1}.- Quantum coherence of the Class $I$ and Class $II$ for $\rho $ are
measured, respectively, by 
\begin{eqnarray}
D_{[A|B]}(\rho )&=&\underset{U_{A},U_{B}}{\min }C_{[A|B]}(\rho ),
\end{eqnarray}
and
\begin{eqnarray}
D_{[B|A]}(\rho )&=&\underset{U_{A},U_{B}}{\min }C_{[B|A]}(\rho ),
\end{eqnarray}%
which describe the minimal quantum coherence with different frames
taken into account.

With this definition, we can arrive at the following conclusion.

\textit{Theorem 1}.- $D_{[A|B]}(\rho )$ and $D_{[B|A]}(\rho )$ are consistent with the geometric
discord of $\rho $ with measurements performed on the corresponding side.

\textit{Proof}. Insert Eq. (2) and Eq. (3) into Eq. (5), we can arrive at%
\begin{eqnarray}
&&D_{[A|B]}(\rho ) =\min \sum\limits_{k^{\prime
},l^{\prime },k\neq l}\left\langle kk^{\prime }\right\vert \rho
_{U_{A}\otimes U_{B}}\left\vert ll^{\prime }\right\rangle \left\langle
ll^{\prime }\right\vert \rho _{U_{A}\otimes U_{B}}\left\vert kk^{\prime
}\right\rangle  \notag \\
&&=\min\left\{\sum\limits_{k^{\prime },l^{\prime },k,l}\left\langle kk^{\prime
}\right\vert \rho _{U_{A}\otimes U_{B}}\left\vert ll^{\prime }\right\rangle
\left\langle ll^{\prime }\right\vert \rho _{U_{A}\otimes U_{B}}\left\vert
kk^{\prime }\right\rangle\right.  \notag \\
&&-\left.\sum\limits_{k^{\prime },l^{\prime },k}\left\langle kk^{\prime
}\right\vert \rho _{U_{A}\otimes U_{B}}\left\vert kl^{\prime }\right\rangle
\left\langle kl^{\prime }\right\vert \rho _{U_{A}\otimes U_{B}}\left\vert
kk^{\prime }\right\rangle \right \} \notag \\
&&=Tr\rho ^{2}-\max_{\tilde{k}}Tr\sum\limits_{\tilde{k}}\left( \left\vert 
\tilde{k}\right\rangle \left\langle \tilde{k}\right\vert \otimes
1_{n}\right) \rho \left( \left\vert \tilde{k}\right\rangle \left\langle 
\tilde{k}\right\vert \otimes 1_{n}\right) \rho  \notag \\
&&=\min_{\tilde{k}}\left\Vert \rho -\sum_{\tilde{k}}\left\vert \tilde{k}%
\right\rangle \left\langle \tilde{k}\right\vert \otimes \rho _{\tilde{k}%
\tilde{k}}\right\Vert ^{2},
\end{eqnarray}%
where $\left\vert \tilde{k}\right\rangle =U_{A}\left\vert k\right\rangle $
and $\rho _{\tilde{k}\tilde{k}}=\left( \left\langle \tilde{k}\right\vert
\otimes 1_{n}\right) \rho \left( \left\vert \tilde{k}\right\rangle \otimes
1_{n}\right) $ with $1_{n}$ the n-dimensional identity. It is obvious that
Eq. (7) is consistent with the definition of the geometric quantum discord
[16], which means that the quantum coherence measure $D_{[A|B]}(\rho )$ is
the geometric quantum discord.

Analogously, the quantum coherence measure
of Class $II$ can be written as 
\begin{eqnarray}
D_{[B|A]}(\rho ) &=&\underset{U_{A},U_{B}}{\min }C_{[B|A]}(\rho )  \notag \\
&=&\min_{\tilde{k}^{\prime }}\left\Vert \rho -\sum_{\tilde{k}}\rho _{\tilde{k%
}^{\prime }\tilde{k}^{\prime }}\otimes \left\vert \tilde{k}^{\prime
}\right\rangle \left\langle \tilde{k}^{\prime }\right\vert \right\Vert ^{2},
\end{eqnarray}%
where $\left\vert \tilde{k}^{\prime }\right\rangle =U_{B}\left\vert
k^{\prime }\right\rangle $ and $\rho _{\tilde{k}^{\prime }\tilde{k}^{\prime
}}=\left( 1_{m}\otimes \left\langle \tilde{k}^{\prime }\right\vert \right)
\rho \left( 1_{m}\otimes \left\vert \tilde{k}^{\prime }\right\rangle \right) 
$ with $1_{m}$ the m-dimensional identity. It is obvious that Eq. (8) is also a geometric discord of the other side.
\hfill$\blacksquare$

Both Eq. (7) and Eq. (8) show that geometric quantum discords actually
quantify the quantum coherence. Since quantum coherence is local-basis
dependent, quantum discord also depends on the local basis, which means it
might be increased by local operations. Based on Ref. [31], one can find
that for bipartite systems of qubits, $D_{[B|A]}(\rho )$ and $D_{[A|B]}(\rho
)$ can be analytically solved. In fact, from the point of calculation of
view, one can find that the direct starting with our definition can lead to
a relatively simple procedure. For example, based on the proof of Theorem 1, Eq. (5) will arrive at
\begin{eqnarray}
&&D_{[A|B]}(\rho )=Tr\rho ^{2}-\max_{\tilde{k}}Tr\sum\limits_{\tilde{k}}\left( \left\vert 
\tilde{k}\right\rangle \left\langle \tilde{k}\right\vert \otimes
1_{n}\right) \rho \left( \left\vert \tilde{k}\right\rangle \left\langle 
\tilde{k}\right\vert \otimes 1_{n}\right) \rho\notag\\
&&=Tr\rho ^{2}-
\max \sum_{\tilde{k},i,j,i^\prime,j^\prime}\frac{1}{16}\left[4+2 x_ix_j\left\langle \tilde{k} \right\vert\sigma_i\left\vert \tilde{k} \right\rangle\left\langle\tilde{k}\right\vert\sigma_j\left\vert\tilde{k}\right\rangle\right.\notag\\
&&+\left.2  y_iy_j Tr\left\{\sigma_i \sigma_j\right\}+T_{ij}T_{i^\prime j^\prime}\left\langle\tilde{k}\right\vert\sigma_i\left\vert\tilde{k}\right\rangle\left\langle\tilde{k}\right\vert\sigma_{i^\prime}\left\vert \tilde{k}\right\rangle Tr \left\{\sigma_j \sigma_{j^\prime}\right\}\right]\notag\\
&&=Tr\rho ^{2}-\frac{1}{4}\left(1+\left\Vert y\right\Vert^2\right)\notag\\
&&-\max \sum_{i,j,i^\prime}\frac{1}{4}\left[x_ix_j+\left.T_{ii^\prime}T_{ji^\prime}\right]\left\langle \psi
_{1}^{1}\right\vert \sigma _{j}\left\vert \psi _{1}^{1}\right\rangle
\left\langle \psi _{1}^{1}\right\vert \sigma _{i}\left\vert \psi
_{1}^{1}\right\rangle \right.\notag\\
&&=\frac{1}{4}(\left\Vert T\right\Vert ^{2}+\left\Vert \vec{x}%
\right\Vert ^{2}-\max \vec{p}_{1}^{t}M\vec{p}_{1}),
\end{eqnarray}
 and similarly,  Eq. (6) will directly
arrive at 
\begin{equation}
D_{[B|A]}(\rho )=\frac{1}{4}\left( \left\Vert \vec{y}\right\Vert
^{2}+\left\Vert T\right\Vert ^{2}-\max \vec{p}_{2}^{t}N\vec{p}_{2})\right) ,
\end{equation}%
where $M=\vec{x}\vec{x}^{t}+TT^{t}$, $N=\vec{y}\vec{y}^{t}+T^{t}T,$ 
and $\vec{x}$, $\vec{y},T$ are the Bloch vectors and tensor obtained from
the Bloch representation of $\rho $, the superscript $t$ means transpose. In addition, $\sigma_k$ in above equations denote the Pauli matrices, $\left\vert\tilde{k}\right\rangle$ is defined the same as that in Eq. (7), $\left\vert\psi_i^j\right\rangle$ denotes the $i$th orthonormal vector of the complete set ${\left\vert\tilde{k}\right\rangle}$ of $j$th subsystem and $\vec{p}_{j}$ is the Bloch vector of $\left\vert\psi_1^1\right\rangle$ . Thus, we can easily find that $D_{[A|B]}(\rho )=%
\frac{1}{4}\left( \left\Vert \vec{x}\right\Vert ^{2}+\left\Vert T\right\Vert
^{2}-\lambda _{1\max }\right) $ and $D_{[B|A]}(\rho )=\frac{1}{4}\left(
\left\Vert \vec{y}\right\Vert ^{2}+\left\Vert T\right\Vert ^{2}-\lambda
_{2\max }\right) $ with $\lambda _{1\max }$, $\lambda _{2\max }$ the maximal
eigenvalue of $\vec{x}\vec{x}^{t}+TT^{t}$ and $\vec{y}\vec{y}^{t}+T^{t}T$,
respectively. These results also provide a demonstration of the consistency
of quantum discord and our coherence measure in two-qubit systems.

\subsection{Symmetric quantum correlation}

We have considered the partial contribution of quantum coherence which shows
the coincidence between geometric quantum discord and quantum coherence
measure. Now we turn to addressing the contribution of all the coherence of
a density matrix $\rho $. A natural method to extracting the coherence is to
collect all the off-diagonal elements by
\begin{equation}
\tilde{C}_{Total}(\rho )=\sum_{\left( kk^{\prime }\right) \neq \left(
ll^{\prime }\right) }\left\vert \left\langle kk^{\prime }\right\vert \rho
_{U_{A}\otimes U_{B}}\left\vert ll^{\prime }\right\rangle \right\vert ^{2}.
\end{equation}%

\textit{Definition 2}.- The measure of the total coherence can be defined as%
\begin{equation}
\tilde{D}(\rho )=\underset{U_{A},U_{B}}{\min }\tilde{C}_{Total}(\rho ).
\end{equation}%
From this definition,  one can arrive at the following rigorous conclusion.

\textit{Theorem 2}.- $\tilde{D}(\rho )$ is consistent with the geometric
discord of $\rho $ with two-side measurements.

\textit{Proof. }%
\begin{eqnarray}
&&\tilde{D}(\rho )  \notag \\
&=&\min \left( \sum_{kk^{\prime }ll^{\prime }}\left\vert \left\langle
kk^{\prime }\right\vert \rho _{U_{A}\otimes U_{B}}\left\vert ll^{\prime
}\right\rangle \right\vert ^{2}-\sum_{\left( kk^{\prime }\right) =\left(
ll^{\prime }\right) }\left\vert \left\langle kk^{\prime }\right\vert \rho
_{U_{A}\otimes U_{B}}\left\vert ll^{\prime }\right\rangle \right\vert
^{2}\right)  \notag \\
&=&\min \left( Tr\rho _{U_{A}\otimes U_{B}}^{2}-\sum_{kl}\left\vert
\left\langle kl\right\vert \rho _{U_{A}\otimes U_{B}}\left\vert
kl\right\rangle \right\vert ^{2}\right)  \notag \\
&=&\min \left\Vert \rho -\sum_{kl}\left( \Pi _{k}\otimes \Pi _{l}\right)
\rho \left( \Pi _{k}\otimes \Pi _{l}\right) \right\Vert ^{2},
\end{eqnarray}%
with $\Pi _{i}=U_{A/B}\left\vert i\right\rangle \left\langle i\right\vert
U_{A/B}^{\dagger }$ the projector on A/B qubit in some basis $%
U_{A/B}\left\vert i\right\rangle $. It is obvious that Eq. (13) is actually
the geometric discord with two-side measurements or the symmetric quantum discord.
\hfill$\blacksquare$

It has been shown that $\tilde{D}(\rho )$ is completely consistent with the geometric discord with
two-side measurements, but it is hard to analytically calculated even for a
general two-qubit mixed state. This can be seen from the recent analysis
given in Ref. [47]. In order to obtain an analytic expression at least for the
two-qubit case, we would like to extract all the coherence with the
following approach: 
\begin{equation}
C_{Total}(\rho )=C_{[A|B]}(\rho )+C_{[B|A]}(\rho ),
\end{equation}%
which means that we have considered the contribution of the doubled
coherence corresponding to the anti-diagonal entries. This is in fact
completely valid from the measure point of view, because this does not
influence the nature of coherence, but the relative value of the coherence.
In addition, We can easily prove that $D(\rho )$ given in the following equation has
an interesting property that will be shown by the corollary in the next subsection. Therefore, we would like to
introduction the below definition.

\textit{Definition 3}.- The total  quantum coherence can also be alternatively defined  by
\begin{equation}
D(\rho )=\underset{U_{A},U_{B}}{\min }C_{Total}(\rho ).
\end{equation}%
Similarly, in this definition, we also consider the  possible minimal value of the contribution of anti-diagonal entries. $D(\rho )$ in this definition 
can be easily calculated by the following theorem.

\textit{Theorem 3} .- The total quantum coherence measure $D(\rho )$ is given
by%
\begin{equation}
D(\rho )=D_{[A|B]}(\rho )+D_{[B|A]}(\rho ),
\end{equation}%
which can grasp the symmetric quantum correlation.

\textit{Proof. } Substitute Eq. (3) and Eq. (4) into Eq. (16), one will obtain
that%
\begin{gather}
D(\rho )=-\max_{\tilde{k}\tilde{k}^{\prime }}\left[ Tr\right. \sum\limits_{%
\tilde{k}\tilde{k}^{\prime }}\left( \left\vert \tilde{k}\right\rangle
\left\langle \tilde{k}\right\vert \otimes 1_{n}\right) \rho \left(
\left\vert \tilde{k}\right\rangle \left\langle \tilde{k}\right\vert \otimes
1_{n}\right) \rho  \notag \\
+Tr\sum\limits_{\tilde{k}^{\prime }}\left( 1_{m}\otimes \left\vert \tilde{k}%
^{\prime }\right\rangle \left\langle \tilde{k}^{\prime }\right\vert \right)
\rho \left( 1_{m}\otimes \left\vert \tilde{k}^{\prime }\right\rangle
\left\langle \tilde{k}^{\prime }\right\vert \right) \left. \rho \right]
+2Tr\rho ^{2}.
\end{gather}%
It is obvious that $\left\vert \tilde{k}\right\rangle $ and $\left\vert 
\tilde{k}^{\prime }\right\rangle $ are independent. Thus $C_{[A|B]}(\rho )$
and $C_{[B|A]}(\rho )$ can be optimized separately. Therefore, we have%
\begin{eqnarray}
D(\rho ) &=&\underset{U_{A},U_{B}}{\min }C_{[A|B]}(\rho )+\underset{%
U_{A},U_{B}}{\min }C_{[B|A]}(\rho ) \\
&=&D_{[A|B]}(\rho )+D_{[B|A]}(\rho ),
\end{eqnarray}%
which happens to be the sum of the two classes of quantum coherence $%
D_{[A|B]}(\rho )$ and $D_{[B|A]}(\rho )$.

Now we prove that $D(\rho )$ can grasp the symmetric quantum correlation.
It is obvious that $D(\rho )$ does not depend on the exchange of $A$ and $B$%
. If $D(\rho )=0$, it means that there exist $U_{A}^{\prime }$ and $%
U_{B}^{\prime }$ such that $\rho _{U_{A}^{\prime }\otimes U_{B}^{\prime }}$
is a diagonal matrix. In the same basis, the reduced matrices are also
diagonal. That is, the density matrix $\rho $ have a diagonal form in the
product of its marginal bases [33,34]. So $\rho _{U_{A}^{\prime }\otimes
U_{B}^{\prime }}$ has no quantum correlation. If a density matrix $\rho $ is
classical correlated, then it must have a diagonal form in the product of
its marginal bases. Therefore, based on our definition of $D(\rho )$, we
must be able to find such $U_{A}^{\prime }\ $and $U_{B}^{\prime }$ that $%
\rho _{U_{A}^{\prime }\otimes U_{B}^{\prime }}$ has no off-diagonal elements
in the basis. That is, $D(\rho )$ vanishes. Hence, $D(\rho )$ does not only
measure the total coherence, but also it can serve as a symmetric measure of
quantum correlation. The proof is completed.\hfill$\blacksquare$

\subsection{Concurrence by the monogamy}

As mentioned above, the introduction of the new measure of the total coherence lies in its intriguing properties.
Now we will first show how this measure is connected with the bipartite concurrence of pure states.
Let $\rho _{AB}=\left\vert \psi \right\rangle _{AB}\left\langle \psi
\right\vert $ be an $(m\otimes n)-$ dimensional bipartite pure state, $\rho
_{A}=$Tr$_{B}\rho _{AB}$ and $\rho _{B}=$Tr$_{B}\rho _{AB}$. Suppose $%
\left\vert \psi \right\rangle _{AB}=\sum_{ij}a_{ij}\left\vert
ij\right\rangle $ in the computational basis, then we can obtain the
concurrence of this state [48,49] is 
\begin{equation}
\mathcal{C}\left( \left\vert \psi \right\rangle _{AB}\right) =2\sqrt{%
\sum_{i<j,k<l}\left\vert a_{ik}a_{jl}-a_{il}a_{jk}\right\vert ^{2}}.
\end{equation}%
Thus one will obtain the following theorem.

\textit{Theorem 4}.-The concurrence of the state $\left\vert \psi
\right\rangle _{AB}$ and the total coherence as well as the local coherence
have the following monogamy relation: 
\begin{equation}
\mathcal{C}^{2}\left( \left\vert \psi \right\rangle _{AB}\right) =\mathcal{D}%
(\rho _{AB})-\mathcal{D}(\rho _{A})-\mathcal{D}(\rho _{B}),
\end{equation}%
with $\mathcal{D}\left( \cdot \right) =C_{Total}(\cdot )$ for convenience.

\textit{Proof.} Based on the definition of $C_{Total}(\rho _{AB})$, we have 
\begin{eqnarray}
\mathcal{D}(\rho _{AB}) &=&C_{[A|B]}(\rho _{AB})+C_{[B|A]}(\rho _{AB}) 
\notag \\
&=&2\sum_{\left( ij\right) <\left( kl\right) }\left\vert a_{ij}a_{kl}^{\ast
}\right\vert ^{2}+2\sum_{i<k,j\neq l}\left\vert a_{ij}a_{kl}^{\ast
}\right\vert ^{2}  \notag \\
&=&2\sum_{k,\left( j<l\right) }\left\vert a_{kj}a_{kl}^{\ast }\right\vert
^{2}+2\sum_{\left( i<k\right) ,j}\left\vert a_{ik}a_{jk}^{\ast }\right\vert
^{2}  \notag \\
&&+4\sum_{i<k,j\neq l}\left\vert a_{ij}a_{kl}^{\ast }\right\vert ^{2},
\end{eqnarray}%
and%
\begin{eqnarray*}
&&\mathcal{D}(\rho _{A})+\mathcal{D}(\rho _{B}) \\
&=&2\sum_{i<j}\left\vert \sum_{k}a_{ik}a_{jk}^{\ast }\right\vert
^{2}+2\sum_{i<j}\left\vert \sum_{k}a_{ki}a_{kj}^{\ast }\right\vert ^{2} \\
&=&2\sum_{\left( i<j\right) ,k}\left\vert a_{ik}a_{jk}^{\ast }\right\vert
^{2}+2\sum_{\left( i<j\right) ,k}\left\vert a_{ki}a_{kj}^{\ast }\right\vert
^{2} \\
&&+2\sum_{i<j,k\neq l}a_{ik}a_{jk}^{\ast }a_{il}^{\ast
}a_{jl}+2\sum_{i<j,k\neq l}a_{ki}a_{kj}^{\ast }a_{li}^{\ast }a_{lj}.
\end{eqnarray*}%
Therefore 
\begin{eqnarray}
&&\mathcal{D}(\rho _{AB})-\mathcal{D}(\rho _{A})-\mathcal{D}(\rho _{B}) 
\notag \\
&=&4\sum_{i<k,j\neq l}\left\vert a_{ij}a_{kl}^{\ast }\right\vert
^{2}-2\sum_{i<j,k\neq l}a_{ik}a_{jk}^{\ast }a_{il}^{\ast }a_{jl}  \notag \\
&&-2\sum_{i<j,k\neq l}a_{ki}a_{kj}^{\ast }a_{li}^{\ast }a_{lj}  \notag \\
&=&\left( 4\sum_{i<j,k<l}a_{ik}a_{jl}^{\ast }a_{ik}^{\ast }a_{jl}\right.
-2\sum_{i<j,k<l}a_{ik}a_{jk}^{\ast }a_{il}^{\ast }a_{jl}  \notag \\
&&\left. -2\sum_{i<j,k<l}a_{ki}a_{kj}^{\ast }a_{li}^{\ast }a_{lj}\right)
+\left( 4\sum_{i<j,k>l}a_{ik}a_{jl}^{\ast }a_{ik}^{\ast }a_{jl}\right. 
\notag \\
&&-2\sum_{i<j,k>l}a_{ik}a_{jk}^{\ast }a_{il}^{\ast }a_{jl}\left.
-2\sum_{i<j,k>l}a_{ki}a_{kj}^{\ast }a_{li}^{\ast }a_{lj}\right)  \notag \\
&=&4\sum_{i<j,k<l}\left\vert a_{ik}a_{jl}-a_{il}a_{jk}\right\vert ^{2},
\end{eqnarray}%
which is just consistent with $\mathcal{C}^{2}\left( \left\vert \psi
\right\rangle _{AB}\right) $. The proof is finished.\hfill$\blacksquare$

From Theorem 4, one can easily find that Eq. (21) has the similar form as the
monogamy relation between the bipartite concurrence and the 3-tangle [35]. It
directly demonstrates the distribution of coherence and provides some limitation on the coherence of the subsystem in some
given basis. It also gives us an alternative understanding of pure-state
entanglement. Actually, Theorem 4 can also be generalized to the mixed
state, which is given by the following corollary.

\textit{Corollary 1}. For a mixed state $\rho _{AB}$, the monogamy relation
given in theorem 3 can become the following inequality,%
\begin{equation}
\mathcal{C}^{2}(\rho _{AB})\leq \min_{\{p_{i},\left\vert \varphi
_{i}\right\rangle _{AB}\}}\sum p_{i}\mathcal{D}(\left\vert \varphi
_{i}\right\rangle _{AB})-\mathcal{D}(\rho _{A})-\mathcal{D}(\rho _{B}).
\end{equation}%
\textit{Proof}. Let $\rho _{AB}=\sum_{i}p_{i}\left\vert \varphi
_{i}\right\rangle _{AB}\left\langle \varphi _{i}\right\vert $ be the
decomposition of $\rho _{AB}$ that achieves the optimal average as $%
\min_{\{p_{i},\left\vert \varphi _{i}\right\rangle _{AB}\}}\sum p_{i}%
\mathcal{D}(\left\vert \varphi _{i}\right\rangle _{AB})$. For each pure
state $\left\vert \varphi _{i}\right\rangle _{AB}$, one can always use
theorem 3 and obtain the corresponding monogamy equation. Add all the
monogamy equations, we will arrive at%
\begin{equation}
\sum p_{i}\mathcal{C}^{2}\left( \left\vert \varphi _{i}\right\rangle
_{AB}\right) =\sum p_{i}\left[ \mathcal{D}(\left\vert \varphi
_{i}\right\rangle _{AB})-\mathcal{D}(\rho _{A}^{i})-\mathcal{D}(\rho
_{B}^{i})\right]
\end{equation}%
with $\rho _{A}^{i}=Tr_{B}\left\vert \varphi _{i}\right\rangle
_{AB}\left\langle \varphi _{i}\right\vert $ and $\rho
_{B}^{i}=Tr_{A}\left\vert \varphi _{i}\right\rangle _{AB}\left\langle
\varphi _{i}\right\vert $. Since $\rho _{A/B}=\sum_{i}p_{i}\rho _{A/B}^{i}$
and $\mathcal{D}(\rho _{A}^{i})=\sum\limits_{j\neq k}\left\vert \left( \rho
_{A}^{i}\right) _{jk}\right\vert ^{2}$, based on the convexity of $\mathcal{D%
}(\rho _{A}^{i})$ we have%
\begin{equation}
\sum p_{i}\mathcal{C}^{2}\left( \left\vert \varphi _{i}\right\rangle
_{AB}\right) \leq \sum p_{i}\mathcal{D}(\left\vert \varphi _{i}\right\rangle
_{AB})-\mathcal{D}(\rho _{A})-\mathcal{D}(\rho _{B}).
\end{equation}%
Because the concurrence $\mathcal{C}\left( \left\vert \varphi
_{i}\right\rangle _{AB}\right) $ is also a convex function, it follows that%
\begin{equation}
\sum p_{i}\mathcal{C}^{2}\left( \left\vert \varphi _{i}\right\rangle
_{AB}\right) \geq \left[ \sum p_{i}\mathcal{C}\left( \left\vert \varphi
_{i}\right\rangle _{AB}\right) \right] ^{2}\geq \mathcal{C}^{2}\left( \rho
_{AB}\right) .
\end{equation}%
Eq. (26) and Eq. (27)  show that the proposition is right, which completes the
proof.
\hfill$\blacksquare$

Again Eq. (24) is similar to that of the monogamy relation for the
mixed state given in Ref. [35]. Eq. (24) shows that the squared concurrence plus
the local coherence will be never larger than the average total coherence.
In fact, from our theorem 3, one can easily find the exact value of
geometric discord and our suggested symmetric quantum correlation measure
without any optimization procedure, which is also one of the reasons
why we consider the doubled anti-diagonal entries in the definition of the
symmetric correlation measure.

\textit{Corollary 2}.- For a bipartite pure state $\rho =\left\vert \phi
\right\rangle \left\langle \phi \right\vert $ defined in arbitrary ($%
m\otimes n$) dimension, the geometric discord and the symmetric quantum
correlations can be given by the concurrence as 
\begin{equation}
D(\rho )=\mathcal{C}^{2}(\rho )
\end{equation}%
and 
\begin{equation}
D_{[A|B]}(\rho )=D_{[B|A]}(\rho )=\frac{1}{2}\mathcal{C}^{2}(\rho ).
\end{equation}

\textit{Proof}. Since $\rho $ is a pure state, theorem 3 holds for $\rho $.
That is, 
\begin{equation}
\mathcal{C}^{2}(\rho )=\mathcal{D}(\rho )-\mathcal{D}(\rho _{A})-\mathcal{D}%
(\rho _{B}),
\end{equation}%
with $\rho _{A}=Tr_{B}\rho $, $\rho _{B}=Tr_{A}\rho $, $\mathcal{D}(\rho
_{A})=\sum_{\substack{ \ i\neq j}}\left\vert \rho _{Aij}\right\vert ^{2}$
and $\mathcal{D}(\rho _{B})=\sum_{\substack{ \ i\neq j}}\left\vert \rho
_{Bij}\right\vert ^{2}$. It is obvious that Eq. (30) is independent of the
basis. So we can select $U_{A}$ and $U_{B}$ such that $U_{A}\rho
_{A}U_{A}^{\dagger }\ $and $U_{B}\rho _{B}U_{B}^{\dagger }$ are diagonal,
thus we have 
\begin{equation}
\mathcal{D}(U_{A}\rho _{A}U_{A}^{\dagger })=\mathcal{D}(U_{B}\rho
_{B}U_{B}^{\dagger })=0.
\end{equation}
It easily turns out 
\begin{equation}
D(\rho _{AB})=\min_{U_{A}\otimes U_{B}}\mathcal{D}(\rho _{AB\left(
U_{A}\otimes U_{B}\right) })\leq \mathcal{C}^{2}(\rho _{AB}).
\end{equation}%
On the contrary, select $U_{A}^{^{\prime }}$ and $U_{B}^{^{\prime }}$ such
that the optimization $\min_{U_{A}\otimes U_{B}}\mathcal{D}(\rho _{AB\left(
U_{A}\otimes U_{B}\right) })\mathcal{\ }$is attained, i.e. 
\begin{equation}
D(\rho _{AB})=\mathcal{D}(\rho _{AB\left( U_{A}^{\prime }\otimes
U_{B}^{\prime }\right) }).
\end{equation}%
However, $\mathcal{D}(\rho _{A})$ and $\mathcal{D}(\rho _{B})$ could be
non-zero, so from Eq. (30), one will arrive at 
\begin{equation}
D(\rho _{AB})\geqslant \mathcal{C}^{2}(\rho _{AB}).
\end{equation}
Eq. (32) and Eq. (34) show $D(\rho _{AB})=\mathcal{C}^{2}(\rho _{AB})$.

Next we will turn to the proof of Eq. (29). Since $D_{[A|B]}(\rho )$ and $%
D_{[B|A]}(\rho )$ are invariant under local unitary operations, we would
like to consider the state $\rho $ after the Schmidt decomposition. In this
case, $\rho $ can be written by 
\begin{equation}
\rho =\sum_{i,j=0}^{\min \{m,n\}}\sigma _{i}\sigma _{j}\left\vert
ii\right\rangle \left\langle jj\right\vert ,
\end{equation}%
with $\sigma _{i}$ the Schmidt coefficients. Based on the definitions of $%
D_{[A|B]}(\rho )$ and $D_{[B|A]}(\rho )$ given in Eq. (3) and Eq. (4),
respectively, that is, 
\begin{equation}
D_{[A|B]}(\rho )=\sum_{\left( \alpha \neq \gamma \right) ,\beta ,\delta
,}\sum_{i,j=0}^{\min \{m,n\}}\left\vert \sigma _{i}\sigma _{j}\left\langle
\alpha \beta \right. \left\vert ii\right\rangle \left\langle jj\right\vert
\left. \gamma \delta \right\rangle \right\vert ^{2}
\end{equation}%
and 
\begin{equation}
D_{[B|A]}(\rho )=\sum_{\left( \beta \neq \delta \right) ,\alpha ,\gamma
,}\sum_{i,j=0}^{\min \{m,n\}}\left\vert \sigma _{i}\sigma _{j}\left\langle
\alpha \beta \right. \left\vert ii\right\rangle \left\langle jj\right\vert
\left. \gamma \delta \right\rangle \right\vert ^{2},
\end{equation}%
where $\left\vert \alpha \right\rangle ,\left\vert \beta \right\rangle
,\left\vert \gamma \right\rangle $ and $\left\vert \delta \right\rangle $
are the optimal local orthonormal basis in their corresponding subspace such
that the optimization of $D_{[A|B]}(\rho )$ and $D_{[B|A]}(\rho )$ is
achieved. From the two Eqs. (36) and (37), one can easily find that $%
D_{[A|B]}(\rho )$ will become $D_{[B|A]}(\rho )$ and \textit{vice versa}, 
if we exchange their subspace. Thus we have \ $D_{[A|B]}(\rho )=$ $%
D_{[B|A]}(\rho )$. According to Eq. (28) and Theorem 2, it is natural that Eq. (29) hold.
The proof is finished.\hfill$\blacksquare$

\subsection{Possible connection with nonlocality: violation of CHSH
inequality for two qubits}

In this part, for the integrity we will quantify the third class of quantum
coherence which corresponds to the overlap of quantum coherence in Class 
\textit{I} and Class \textit{II}. From our classification, we can find that
the quantum coherence of Class \textit{III} does not depend on the exchange
of subsystems A and B. In the same way as the quantification of quantum
coherence of Class \textit{I} and Class \textit{II}, we can extract all the
anti-diagonal entries of $\rho $ within all possible local frames as 
\begin{equation}
\tilde{\Delta}_{kk^{\prime },ll^{\prime }}=\left\langle kk^{\prime
}\right\vert \rho _{U_{A}\otimes U_{B}}\left\vert ll^{\prime }\right\rangle
,\left\{ 
\begin{array}{c}
k+l=m-1 \\ 
k^{\prime }+l^{\prime }=n-1%
\end{array}%
\right. .
\end{equation}%
Thus the contribution of anti-diagonal entries can be described as 
\begin{equation}
v(\rho )=\sum\limits_{\substack{ k+l=m-1  \\ k^{\prime }+l^{\prime }=n-1}}%
\left\vert \tilde{\Delta}_{kk^{\prime },ll^{\prime }}\right\vert ^{2}.
\end{equation}%
Similarly to the previous definitions, considering all the potential $U_{A}$ and $U_{B}$, we will arrive at a new definition.

\textit{Definition 4}.- The third class of coherence can be measured by 
\begin{equation}
V(\rho )=\underset{U_{A},U_{B}}{\min }v(\rho ).
\end{equation}%
A dual measure of coherence can also be defined as
\begin{equation}
\tilde{V}(\rho )=\underset{U_{A},U_{B}}{\max }v(\rho ).
\end{equation}%

In fact, what $V(\rho )$ and $\tilde{V}(\rho )$ characterize for a general
state except the coherence defined by us has  been unknown yet. This is also
why we consider the contribution of the anti-diagonal entries by
introducing both the maximum and the minimum. Of course, a simple result can be seen for the pure states of two qubits. That is,  $V(\rho )=0$ means that the two-qubit pure state is separable. In fact, for the system of two qubits, one can further find that both $V(\rho )$ and $\tilde{V}(\rho )$ are closely
related to the violation of the remarkable CHSH inequality. In this sense,
we would like to make the following conjecture.

\textit{Conjecture}.- At least one of $V(\rho )$ and $\tilde{V}%
(\rho )$ could be related to the nonlocality subject to some Bell
theory.

 Next we will show how $V(\rho )$ and $\tilde{V}(\rho )$ are
connected with the violation of CHSH inequality for two qubits. Substitute the Bloch
representation of $\rho _{AB}$ into Eq. (39), it follows that
\begin{eqnarray}
&&v(\rho )=\frac{1}{4}\sum\limits_{i,j=1}^{3}T_{ij}T_{ij}-\frac{1}{4}%
\sum\limits_{i,k,l=1}^{3}T_{ik}T_{il}\left\langle \psi _{1}^{2}\right\vert
\sigma _{k}\left\vert \psi _{1}^{2}\right\rangle  \notag \\
&&\times \left\langle \psi _{1}^{2}\right\vert \sigma _{l}\left\vert \psi
_{1}^{2}\right\rangle -\frac{1}{4}\sum\limits_{i,j,k=1}^{3}T_{ik}T_{jk}\left%
\langle \psi _{1}^{1}\right\vert \sigma _{i}\left\vert \psi
_{1}^{1}\right\rangle \left\langle \psi _{1}^{1}\right\vert \sigma
_{j}\left\vert \psi _{1}^{1}\right\rangle  \notag \\
&&+\frac{1}{4}\sum\limits_{i,j,k,l=1}^{3}T_{ik}T_{jl}\left\langle \psi
_{1}^{1}\right\vert \sigma _{j}\left\vert \psi _{1}^{1}\right\rangle
\left\langle \psi _{1}^{1}\right\vert \sigma _{i}\left\vert \psi
_{1}^{1}\right\rangle  \notag \\
&&\times \left\langle \psi _{1}^{2}\right\vert \sigma _{k}\left\vert \psi
_{1}^{2}\right\rangle \left\langle \psi _{1}^{2}\right\vert \sigma
_{l}\left\vert \psi _{1}^{2}\right\rangle  \notag \\
&=&\frac{1}{4}(\left\Vert T\right\Vert ^{2}-\vec{p}_{1}^{t}TT^{t}\vec{p}_{1}-%
\vec{p}_{2}^{t}T^{t}T\vec{p}_{2}+\vec{p}_{1}^{t}T\vec{p}_{2}\vec{p}%
_{2}^{t}T^{t}\vec{p}_{1}),
\end{eqnarray}%
where $\left\vert \psi _{i}^{j}\right\rangle $ similar to that in Eq. (9) means the $i$th orthonormal vector subject to $j$th subsystem, $\sigma_i$ denote the Pauli matrices and $\vec{p}%
_{j}$ is the Bloch vector of $\left\vert \psi _{1}^{j}\right\rangle $. Thus
we can obtain the rigorous expressions as follows.

\textit{Theorem 5.}-For a bipartite state of qubits, 
\begin{equation}
V(\rho )=\frac{1}{4}\sigma _{\min }
\end{equation}%
and%
\begin{equation}
\tilde{V}(\rho )=\frac{1}{4}(\left\Vert T\right\Vert ^{2}-\sigma _{\min }),
\end{equation}%
where $\sigma _{\min }$ is the minimal eigenvalue of $TT^{t}$.

\textit{Proof. } In order to give an analytic optimization of Eq. (42), we
would like to turn to a simple basis, since Eq. (42) is not changed under
local unitary transformations. Consider the singular value decomposition of $%
T$ as $T=U\Lambda V$, then $U$ and $V$ are orthogonal matrix and $\Lambda
=diag[\sigma _{1},\sigma _{2},\sigma _{3}]$ with 
\begin{equation}
\sigma _{1}\geq \sigma _{2}\geq \sigma _{3}\geq 0,
\end{equation}
since $T$ is a real matrix. Thus $U$ and $V$ that could appear in Eq. (42) can be absorbed by $%
\vec{p}_{1}$ and $\vec{p}_{2}$ and for the habit, we let $\vec{p}%
_{1}=[x_{1},x_{2},x_{3}]^{t}$ and $\vec{p}_{2}=[y_{1},y_{2},y_{3}]^{t}$
which should be distinguished from the expressions given in Eq. (9) and Eq.
(10). Thus the optimizations defined in Eq. (40) and Eq. (41) are changed into%
\begin{eqnarray}
V(\rho ) &=&\underset{U_{A},U_{B}}{\min }v(\rho )=\frac{1}{4}(\left\Vert
T\right\Vert ^{2}-\max L) \\
\tilde{V}(\rho ) &=&\max_{U_{A},U_{B}}v(\rho )=\frac{1}{4}(\left\Vert
T\right\Vert ^{2}-\min L)
\end{eqnarray}%
with%
\begin{equation}
L(x_{i},y_{i})=\sum_{i=1}^{3}\sigma _{i}^{2}\left(
x_{i}^{2}+y_{i}^{2}\right) -\left( \sum_{i=1}^{3}\sigma
_{i}x_{i}y_{i}\right) ^{2}.
\end{equation}%
Now we would like to first calculate the minimum of $L$. Based on
Cauchy-Schwarz inequality, we can easily find that

\begin{eqnarray}
L &\geq &\sum_{i=1}^{3}\sigma _{i}^{2}\left( x_{i}^{2}+y_{i}^{2}\right)
-\sum_{i=1}^{3}\sigma _{i}^{2}x_{i}^{2}  \notag \\
&=&\sum_{i=1}^{3}\sigma _{i}^{2}y_{i}^{2}\geq \sigma _{3}^{2}=\sigma _{\min
}.
\end{eqnarray}%
It is obvious that the inequality (49) can be saturated if $\vec{p}_{1}=\vec{p}%
_{2}=[0,0,1]^{t}$ based on Eq. (45). So we have that Eq. (44) is satisfied.

 Now we prove Eq. (43). Based on the Lagrange multiplier method, the
Lagrange function of Eq. (48) can be given by%
\begin{eqnarray}
\Phi (x_{i},y_{i},\lambda ,\mu ) &=&L(x_{i},y_{i})+\lambda \left(
\sum_{i=1}^{3}x_{i}^{2}-1\right)  \notag \\
&&+\mu \left( \sum_{i=1}^{3}y_{i}^{2}-1\right) ,
\end{eqnarray}%
with $\lambda ,\mu $ the Lagrange multipliers. Derivatives on the parameters
of $\Phi (x_{i},y_{i},\lambda ,\mu )$ can be given by%
\begin{equation}
\left\{ 
\begin{array}{c}
\frac{\partial \Phi }{\partial x_{i}}=2\left( \sigma _{i}^{2}+\lambda
\right) x_{i}+2A\sigma _{i}y_{i}=0 \\ 
\frac{\partial \Phi }{\partial y_{i}}=2\left( \sigma _{i}^{2}+\mu \right)
y_{i}+2A\sigma _{i}x_{i}=0%
\end{array}%
\right. ,
\end{equation}%
with 
\begin{equation}
A=\sum_{i=1}^{3}\sigma _{i}x_{i}y_{i}.
\end{equation}
The Eq. (51) has non-zero solution requires that there exists at least an $i$
such that 
\begin{equation}
\det \left[ \left( 
\begin{array}{cc}
\sigma _{i}^{2}+\lambda & A\sigma _{i} \\ 
A\sigma _{i} & \sigma _{i}^{2}+\mu%
\end{array}%
\right) \right] =0.
\end{equation}%
First of all, we suppose $\sigma _{1}>\sigma _{2}>\sigma _{3}>0$. It is not
difficult to find that if there exists a single $i=k$ such that Eq. (53)
holds, one can easily find $x_{k}=y_{k}=1,$ and the others are zero. The
extremes of $L$ in this case are $\sigma _{k}^{2}$. If Eq. (53) holds for all $%
i=1,2,3$, one will find that Eq. (51) has no solution. So the remaining is
that two equations in Eq. (53) hold, and one does not hold. Satisfying this
condition, there exists three possibilities. However, it proves that the
procedure of the calculations are similar and their solutions also have the
similar form. Without loss of generality, we set Eq. (53) is only satisfied
for $i=1,2$, which directly implies 
\begin{equation}
x_{3}=y_{3}=0.
\end{equation}%
In particular, we can get an equation group from Eq. (53) as%
\begin{equation}
\left\{ 
\begin{array}{c}
\left( \sigma _{1}^{2}+\lambda \right) \left( \sigma _{1}^{2}+\mu \right)
-\sigma _{1}^{2}A^{2}=0 \\ 
\left( \sigma _{2}^{2}+\lambda \right) \left( \sigma _{2}^{2}+\mu \right)
-\sigma _{2}^{2}A^{2}=0%
\end{array}%
\right. .
\end{equation}%
A direct simplification will lead to%
\begin{equation}
\lambda \mu =\sigma _{1}^{2}\sigma _{2}^{2},
\end{equation}%
and%
\begin{equation}
A^{2}=\frac{\left( \sigma _{1}^{2}+\lambda \right) \left( \sigma
_{1}^{2}+\mu \right) }{\sigma _{1}^{2}}.
\end{equation}%
In addition, based on Eq. (51), one can also find that 
\begin{equation}
y_{i}=\sqrt{\frac{\sigma _{i}+\lambda }{\sigma _{i}+\mu }}x_{i},i=1,2.
\end{equation}%
Substitute Eqs. (56-58) into Eq. (51), we will obtain that%
\begin{equation}
x_{1}^{2}=\frac{\left( \sigma _{2}^{2}+\lambda \right) \sigma _{1}^{2}}{%
\lambda (\sigma _{1}^{2}-\sigma _{2}^{2})}.
\end{equation}%
Insert Eqs. (56-59) into Eq. (48), $L$ can be rewritten as%
\begin{eqnarray}
L_{12} &=&\sigma _{1}^{2}\left( x_{1}^{2}+y_{1}^{2}\right) +\sigma
_{2}^{2}\left( x_{2}^{2}+y_{2}^{2}\right) -A^{2}  \notag \\
&=&2\sigma _{2}^{2}+\left( \sigma _{1}^{2}-\sigma _{2}^{2}\right) \left(
x_{1}^{2}+y_{1}^{2}\right)  \notag \\
&&-\frac{\left( \sigma _{1}^{2}+\lambda \right) \left( \sigma _{1}^{2}+\mu
\right) }{\sigma _{1}^{2}}  \notag \\
&=&\sigma _{2}^{2}-\sigma _{1}^{2}-\frac{\lambda ^{2}+\sigma _{1}^{2}\sigma
_{2}^{2}}{\lambda }  \notag \\
&&+\frac{\left[ \sigma _{1}^{2}\left( \lambda +\sigma _{2}^{2}\right)
+\lambda \left( \lambda +\sigma _{1}^{2}\right) \right] \sigma
_{1}^{2}(\lambda +\sigma _{2}^{2})}{\sigma _{1}^{2}\lambda (\lambda +\sigma
_{2}^{2})}  \notag \\
&=&\sigma _{1}^{2}+\sigma _{2}^{2}\text{,}
\end{eqnarray}%
which provides an extreme. Similarly, if we assume Eq. (53) holds for any $%
i=k,l$, one will always obtain 
\begin{equation}
L_{kl}=\sigma _{k}^{2}+\sigma _{l}^{2}.
\end{equation}%
Actually, one can easily check that the cases that there exist some "$=$"
hold in $\sigma _{1}\geq \sigma _{2}\geq \sigma _{3}$, are completely
covered by the above calculation. We won't repeat the similar calculations.
Compare the solutions given in Eqs. (60,61), one will obtain that the maximum is
obviously provided by Eq. (60). Substitute this solution into Eq. (46), we
arrive at Eq. (43). The proof is completed.\hfill$\blacksquare$

From this theorem, we can directly find the connection with the violation of
CHSH inequality.

\textit{Corollary 3}.-If $\rho $ is a bipartite quantum state of qubits, 
\begin{equation}
\tilde{V}(\rho )>\frac{1}{4}
\end{equation}
or%
\begin{equation}
V(\rho )<\frac{1}{4}\left( \left\Vert T\right\Vert ^{2}-1\right)
\end{equation}
is equivalent to the violation of CHSH  inequality.

\textit{Proof.} From Eq. (47), it follows that $4\tilde{V}(\rho )=\sigma
_{1}^{2}+\sigma _{2}^{2}$, where $\sigma _{i}^{2}$ denote the eigenvalues of $%
TT^{t} $ in decreasing order. In Ref. [50], it was explicitly reported that
if $\sigma _{1}^{2}+\sigma _{2}^{2}>1,$ the state will violate the CHSH
inequality. So in our case, one can easily conclude that if $\tilde{V}(\rho
)>\frac{1}{4}$, the state $\rho $ will violate CHSH inequality. In addition,
it is obvious that 
\begin{equation}
V(\rho )+\tilde{V}(\rho )=\frac{1}{4}\left\Vert T\right\Vert ^{2},
\end{equation}
from which one can find that $\rho $ violate CHSH inequality means $V(\rho )<%
\frac{1}{4}\left( \left\Vert T\right\Vert ^{2}-1\right) $. This is the end
of the proof.

Before the end of this section, we would like to emphasize that, even though
we have found a very interesting connection between the violation of CHSH
inequality and some special quantum coherence (anti-diagonal entries of
density matrices), we have not yet made sure whether this kind of connection
is suitable for the high dimensional bipartite quantum states.

\section{Conclusion and discussion}

We have shown the complete coincidence between geometric quantum discords
and different classes of quantum coherence in terms of the classification of
quantum coherence. The coincidence provides an alternative understanding of
the asymmetric and symmetric geometric quantum discords based on the
fundamental quantum mechanical feature----- quantum coherence. Furthermore,
a recommended total quantum coherence measure has led to a new symmetric
geometric quantum discord. It has been shown that this kind of total quantum
coherence can be used to construct the monogamy relation between the quantum
coherence and quantum concurrence which can also been understood as some
interpretation of the origin of entanglement. What's more, it is also shown
that the coherence of Class $III$ can be directly connected with the
violation of CHSH inequality for two qubits.

In fact, there are many interesting questions for the future. 1) Is the
quantum coherence of Class \textit{III}  connected with the violation of
CHSH inequality in high dimension or some special nonlocality? 2) Replacing
the minimization in Eq. (5), Eq. (6) , Eq. (12) and Eq. (16) by the maximization ones
similar to Eq. (44), one can obtain a different definition of quantum
coherence measure. Can the new definition be connected with other
interesting issues, such as entanglement, nonlocality, correlations and so
on? 3) How can we quantify the quantum coherence of multipartite quantum
states in the same manner and do the different quantum coherence measure
correspond to different multipartite geometric quantum discords either?

\section{Acknowledgements}

Yu thanks Sandu Popescu, Nicolas Brunner, Paul Skrzypczyk, Ralph Francisco
Silva and J. S. Jin for valuable discussion.

\end{document}